\documentstyle[12pt,epsf]{article}

\begin{document}

\baselineskip=18.8pt plus 0.2pt minus 0.1pt

%%%%%%%%%%% Private Macros %%%%%%%%%%%%%
\makeatletter

%\@addtoreset{equation}{section}
%\renewcommand{\theequation}{\thesection.\arabic{equation}}
\renewcommand{\thefootnote}{\fnsymbol{footnote}}
\newcommand{\beq}{\begin{equation}}
\newcommand{\eeq}{\end{equation}}
\newcommand{\bea}{\begin{eqnarray}}
\newcommand{\eea}{\end{eqnarray}}
\newcommand{\nn}{\nonumber \\}
\newcommand{\hs}[1]{\hspace{#1}}
\newcommand{\vs}[1]{\vspace{#1}}
\newcommand{\Half}{\frac{1}{2}}
\newcommand{\p}{\partial}
\newcommand{\ol}{\overline}
\newcommand{\wt}[1]{\widetilde{#1}}
\newcommand{\ap}{\alpha'}
\newcommand{\bra}[1]{\left\langle  #1 \right\vert }
\newcommand{\ket}[1]{\left\vert #1 \right\rangle }
\newcommand{\vev}[1]{\left\langle  #1 \right\rangle }
\newcommand{\vac}{\ket{0}}

\newcommand{\ul}[1]{\underline{#1}}

\makeatother
%%%%%%%%% End of private macros %%%%%%%%%%%

\begin{titlepage}
\title{
\hfill\parbox{4cm}
{\normalsize MIT-CTP-3706\\{\tt hep-th/0511167}}\\
\vspace{1cm}
Fundamental String Solutions in Open String Field Theories
}
\author{Yoji Michishita
\thanks{
{\tt michishi@lns.mit.edu}
}
\\[7pt]
{\it Center for Theoretical Physics}\\
{\it Massachusetts Institute of Technology}\\
{\it Cambridge MA 02142 USA}
}

\date{\normalsize November, 2005}
\maketitle
\thispagestyle{empty}

\begin{abstract}
\normalsize
In Witten's open cubic bosonic string field theory and Berkovits' 
superstring field theory we investigate solutions of the equations of motion
with appropriate source terms, which correspond to Callan-Maldacena solution
in Born-Infeld theory representing fundamental strings ending on
the D-branes. The solutions are given in order by order manner,
and we show some full order properties in the sense of $\ap$-expansion.
In superstring case we show that the solution is 1/2 BPS in full order.
\end{abstract}

%PACS codes: 11.25.-w

%Keywords: string field theory, D-brane,

\end{titlepage}

%%%%%%%%%%%%%%%%%%%%%%%%%%%%%%%%%%%%%%%%%%%%%%%%%%%%%%%%%%%%%%%%%%%
%%%%%%%%%%%%%%%%%%%%%%%%%%%%%%%%%%%%%%%%%%%%%%%%%%%%%%%%%%%%%%%%%%%
\section{Introduction and Summary}

In Witten's cubic open string field theory\cite{w} and its extension to
superstring such as Berkovits' superstring field theory\cite{b},
it is very difficult to construct solutions with coordinate dependence.
This is because string field theory is nonlocal and contains infinitely
many derivatives. It prevents us from investigating behavior of higher
modes and full order properties. (We consider only classical theory and
do not consider string loop correction. Therefore throughout this paper
``full order'' means exactness in the sense of $\ap$-expansion.)

In this paper we investigate an example of such solutions of the equations
of motion with appropriate source terms, of which we can
derive some full order properties: string field theory counterpart of 
Callan-Maldacena solution\cite{cm}. (For a related topic see \cite{bmsst}.)
This solution represents configuration of fundamental strings
emanating from the D-brane. Since it is also a solution of free U(1) gauge
theory, we expect that we can construct the string field theory solutions in
order by order manner, starting from the linearized equation and
introducing higher order source terms.
In section 2 we construct the solution in Witten's string field theory and
see that it has the following properties:
\begin{itemize}
\item The coefficient of the massless component $A_\mu$ is equal to the
 gauge field $\wt{A}_\mu$ in the effective action with full order correction
 in $\ap$.
\item The solution has no tachyon component, and the massless component has no
 higher order correction.
\item Although we have no proof, we give a convincing argument that 
 massive modes have no singularity unlike the massless component.
\item Energy-momentum tensor given in \cite{sen} has no contribution from
 massive modes, and is equal to that of free U(1) gauge theory.
\end{itemize}
In section 4 we construct the solution in Berkovits' superstring field theory
and see that it has almost the same properties as the bosonic one.
Moreover we show that it is 1/2 BPS in full order.

%%%%%%%%%%%%%%%%%%%%%%%%%%%%%%%%%%%%%%%%%%%%%%%%%%%%%%%%%%%%%%%%%%%
%%%%%%%%%%%%%%%%%%%%%%%%%%%%%%%%%%%%%%%%%%%%%%%%%%%%%%%%%%%%%%%%%%%
\section{Solution in Bosonic String Field Theory}

Let us consider one single D$p$-brane in the flat space.
The bosonic quadratic part of its effective action,
in both bosonic and superstring theory, is given by free U(1) gauge theory.
Spacetime-filling D-brane action has only gauge field $\wt{A}_\mu$, and 
lower dimensional D-brane actions are obtained from it by dropping 
dependence on coordinates perpendicular to the D-branes.
We separate spacetime coordinates $x^\mu$ into 
$x^\pm=\frac{1}{\sqrt{2}}(x^0\pm x^1)$, $x^i$ and $x^I$,
where $x^0$ and $x^i$ are directions along the 
D$p$-brane, and $x^1$ and $x^I$ are directions perpendicular to the 
D$p$-brane. Then $\wt{A}_1$ and $\wt{A}_I$ are scalar fields corresponding
to $x^1$ and $x^I$ respectively.

Suppose $\wt{A}_-=0, \wt{A}_i=0, \wt{A}_I=0$,
and $\wt{A}_+=\wt{A}_+(x^i)$, then the linearized equation of motion is
\beq
\sum_i\p_i\p_i\wt{A}_+=0.
\label{laplace}
\eeq
This is Laplace equation, and ``point charge'' configurations give
solutions:
\beq
\wt{A}_+=\sum_n \frac{c_n}{[\sum_i(x^i-x_n^i)^2]^{\frac{p-2}{2}}},
\label{laplacesol}
\eeq
where $c_n$ and $x_n^i$ are constants.
We assumed $p\geq 3$. For $p=1$ solutions are sums of segments of 
linear functions and for $p=2$ sums of $\log\sum_i(x^i-x_n^i)^2$.
In these cases momentum expressions (i.e. Fourier transformations)
of these solutions require introducing infrared regulators.
Since in string field theory we use momentum expression,
we do not consider $p=1$ and $2$ in this paper. 

For this solution the right hand side of (\ref{laplace})
is not actually zero, but a sum of delta function sources.
In \cite{cm} it has been shown that this configuration represents fundamental
strings stretching along $x^1$ direction
and ending on the D-brane at $x^i=x_n^i$, and extension of this solution to
Born-Infeld theory is again given by (\ref{laplacesol}), without corrections. 
In this interpretation the presence of the delta function sources is not a
problem, because the points $x^i=x_n^i$ are not on the worldvolume of the
D-brane (or are regarded to be infinitely far away).

Furthermore in superstring theory this solution is 1/2 supersymmetric,
both in linearized U(1) gauge theory \cite{cm} and Born-Infeld theory
\cite{lpt}.

In fact this solution is an $\ap$-exact solution as shown in \cite{t}
by computing beta function of the worldsheet sigma model.

Since leading order terms of string field theory action give free
U(1) gauge theory, we expect that 
starting from the solution of (\ref{laplace}) we can construct corresponding
solutions of string field equation ``order by order''. In this section
we investigate the solution in Witten's cubic bosonic string field theory.

In the bosonic string field theory the equation of motion is
\beq
Q\Phi+\Phi^2=0.
\eeq
Of course the right hand
side is not actually zero. To get a right solution 
we have to put a source term which we will call $\Delta_n$.

The solution is constructed by expanding $\Phi$ in some
parameter $g$:
\beq
\Phi=g\Phi_0+g^2\Phi_1+g^3\Phi_2+\dots. \label{phiexp}
\eeq
The equation of motion is decomposed into contributions from each order in $g$:
\bea
\Delta_0 & = & Q\Phi_0, \label{beq0} \\
\Delta_1 & = & Q\Phi_1+\Phi_0^2, \label{beq1} \\
\Delta_2 & = & Q\Phi_2+\Phi_0\Phi_1+\Phi_1\Phi_0, \\
 & \vdots & \nn
\Delta_n & = & Q\Phi_n+\sum_{m=0}^{n-1} \Phi_m\Phi_{n-m-1}, \\
 & \vdots & \nonumber
\eea
Massless part of the lowest order equation (\ref{beq0}) is equivalent to that
of free U(1) gauge theory with source terms.
So we take the following $\Phi_0$ which corresponds to (\ref{laplacesol}):
\beq
\Phi_0=\int\frac{d^pk}{(2\pi)^p}A_+(k_i)c\p X^+e^{ik_iX^i},
\eeq
where coordinate expression of $A_+(k_i)$ which is given by
$A_+(x^i)=\int\frac{d^pk}{(2\pi)^p}A_+(k_i)e^{ik_ix^i}$
satisfies Laplace equation with delta function source terms.
Then the string field source term $\Delta_0$ is
\beq
\Delta_0=-\ap\int\frac{d^pk}{(2\pi)^p}k^2A_+(k_i)c\p c\p X^+e^{ik_iX^i}.
\eeq
$\Phi_0$ satisfies Siegel gauge condition: $b_0\Phi_0=0$.
We require that at each order this condition is satisfied: $b_0\Phi_n=0$.
In addition we require that $\Delta_n$ with $n\geq 1$ also satisfy this
condition: $b_0\Delta_n=0$. $\Delta_0$ does not satisfy it.
This means that $\Delta_0$ is the only source for physical components, and
$\Delta_n$ with $n\geq 1$ are for unphysical components. This is desirable
because, when we eliminate all unphysical massive modes by a gauge fixing
condition and solve all equations for physical massive modes, we have to
obtain equation of motion for massless modes with a simple source term
to have a solution corresponding to Callan-Maldacena solution.

By acting $b_0$ to the equations of motion and noticing that
$b_0Q\Phi_n=L_0\Phi_n$, we obtain
\bea
\Phi_1 & = & -\frac{b_0}{L_0}(\Phi_0^2), \\
\Phi_2 & = & -\frac{b_0}{L_0}(\Phi_0\Phi_1+\Phi_1\Phi_0) \nn
 & = & \frac{b_0}{L_0}\left(\Phi_0\frac{b_0}{L_0}(\Phi_0^2)
 +\frac{b_0}{L_0}(\Phi_0^2)\Phi_0\right), \\
 & \vdots & \nn
\Phi_n & = & -\frac{b_0}{L_0}\sum_{m=0}^{n-1} \Phi_m\Phi_{n-m-1}, \\
 & \vdots & \nonumber
\eea
In this manner $\Phi_n$ can be expressed by $\frac{b_0}{L_0}$ and 
$(n+1)$ copies of $\Phi_0$. Since $b_0$ projects out some components of string
fields, we have to check if there is more information extracted from 
the equations of motion by plugging the above solution back into them:
\bea
\Delta_n & = & Q\Phi_n+\sum_{m=0}^{n-1} \Phi_m\Phi_{n-m-1} \nn
 & = & -Q\frac{b_0}{L_0}\sum_{m=0}^{n-1} \Phi_m\Phi_{n-m-1}
 +\sum_{m=0}^{n-1} \Phi_m\Phi_{n-m-1} \nn
 & = & \frac{b_0}{L_0}Q\sum_{m=0}^{n-1} \Phi_m\Phi_{n-m-1}.
\eea
This should be regarded as determining $\Delta_n$ by lower order solutions.
Notice that if lower order $\Phi_m$ in the right hand side of the above
equation satisfy equations of motion without lower order source terms,
then $\Delta_n$ vanishes:
\bea
\Delta_n & = & \frac{b_0}{L_0}\sum_{m=0}^{n-1}
((Q\Phi_m)\Phi_{n-m-1}-\Phi_m(Q\Phi_{n-m-1})) \nn
& = & -\frac{b_0}{L_0}
\left(\sum_{m=1}^{n-1}\sum_{l=0}^{m-1}\Phi_l\Phi_{m-l-1}\Phi_{n-m-1}
-\sum_{m=0}^{n-2}\sum_{l=0}^{n-m-2}\Phi_m\Phi_l\Phi_{n-m-l-2}\right) \nn
& = & -\frac{b_0}{L_0}\left(
\sum_{m=0}^{n-1}\sum_{l=0}^{m-1}\Phi_l\Phi_{m-l-1}\Phi_{n-m-1}
-\sum_{l=0}^{n-1}\sum_{m=l+1}^{n-1}\Phi_l\Phi_{m-l-1}\Phi_{n-m-1}\right) \nn
& = & 0.
\eea
For our solution source terms should not be zero, and we obtain
\beq
\Delta_n=\frac{b_0}{L_0}\sum_{m=0}^{n-1}[\Delta_m, \Phi_{n-m-1}],
\eeq
which means that higher order source terms are induced by lower order ones.

Obviously $\Phi_n$ has no dependence on momenta along $x^\pm$. In addition,
$\Phi_n$ has the following property: Let $\omega$ be any of the vertex 
operators (or Fock space states) which $\Phi_n$ consists of.
Then $X^I$ part of $\omega$ is a Virasoro descendant of the unit
operator i.e. a state constructed by acting $L'_{-n}$ $(n\geq 2)$
on $\ket{0}$, where $L'_{-n}$ are Virasoro operators of $X^I$ part. Moreover,
$n_+(\omega)-n_-(\omega)=n+1$, where
$n_+$ is the number of $\p^mX^+$ (or $\alpha^+_{-m}$) in $\omega$, and 
$n_-$ is the number of $\p^mX^-$ (or $\alpha^-_{-m}$) in $\omega$.

In summary, matter part of $\omega$ is in the following form:
\beq
\prod_{l=1}^{n_+}\alpha^+_{-p_l}\prod_{l=1}^{n_-}\alpha^-_{-q_l}
\prod_l L'_{-t_l}\prod_l\alpha^{i_l}_{-u_l}\ket{k^i}
\quad (n_+-n_-=n+1,\; p_l,q_l,u_l\geq 1,\; t_l\geq 2).
\label{bstates}
\eeq
The structure of $X^I$ part represents symmetry in $X^I$ directions.

This can be proven by induction as follows.
For $n=0$ this is obvious. Suppose $n>0$. 
We take orthonormal basis of the Fock space $\{\ket{\phi_r}\}$ and its
conjugate $\{\bra{\phi^c_r}\}$. These satisfy 
$\vev{\phi^c_s|\phi_r}=\delta_{rs}$.
Corresponding vertex operators are denoted
by $\phi_r$ and $\phi^c_r$ respectively. Coefficient of $\ket{\phi_r}$ in the 
expansion of $\Phi_n$ by $\{\ket{\phi_r}\}$ is given by 
$\vev{\phi^c_r|\Phi_n}$:
\bea
\vev{\phi^c_r|\Phi_n} & = & \vev{\phi^c_r\Bigg|
 -\frac{b_0}{L_0}\sum_{m=0}^{n-1} \Phi_m\Phi_{n-m-1}} \nn
 & = & -\vev{\frac{b_0}{L_0}\phi^c_r\Bigg|
 \sum_{m=0}^{n-1} \Phi_m\Phi_{n-m-1}}.
\label{phincoeff}
\eea
First we concentrate on $X^\pm$ sector. Since $b_0$ affects
only on ghost part and $L_0$ gives a numerical factor for each level,
we can neglect $\frac{b_0}{L_0}$.
By the assumption of the induction, $n_+(\Phi_m)-n_-(\Phi_m)$ is $m+1$ and 
$n_+(\Phi_{n-m-1})-n_-(\Phi_{n-m-1})$ is $n-m$.
There are two processes which change the number of $X^+$ and $X^-$:
contraction and conformal transformations in the star product.
Since $X^+$ has nonzero contraction only with $X^-$ and vice versa,
Both processes preserve the difference of these numbers, and
the total number of $X^+$ and $X^-$ in the correlator should be equal
for nonzero contribution.
Therefore $n_+(\phi_r^c)-n_-(\phi_r^c)$ should be $-n-1$. This means that
$n_+(\phi_r)-n_-(\phi_r)$ is $n+1$.

Next we consider $X^I$ sector. By the assumption of the induction, 
both $\Phi_m$ and $\Phi_{n-m-1}$ are Virasoro descendants of the unit
operator. If $\phi_r^c$ is a descendant of a nontrivial primary field
$\lambda$, by using the well-known procedure relating a correlator with
worldsheet energy-momentum tensors to ones without it,
the correlator reduces to one point function of $\lambda$, which vanishes
because of its nonzero conformal dimension. This means that $\phi_r$
consists of Virasoro descendants of the unit operator.

Ghost part of $\Phi_n$ can also be restricted further as is explained in
\cite{z}.

An immediate consequence of the above fact on the number of $X^\pm$ is
that each coefficient of 
Fock space state in the solution $\Phi$ receives contribution from
only one $\Phi_n$. (Here we regard states consisting of the same oscillators 
with different spacetime indices as different states.)
In particular, the coefficient of the massless vertex operator
$c\p X^\mu e^{ik_\mu X^\mu}$, which is denoted by $A_\mu$,
is never corrected by higher order contribution, and the coefficient 
of the lowest state, which represents tachyon, is zero in full order.
In addition, we see that the inverses of $L_0$ in the expression of 
$\Phi_n$ with $n\geq 1$ do not cause any problem,
because only massless and tachyon components, which is absent in
$\Phi_n$ with $n\geq 1$, are problematic.

We can easily see that $\Delta_n$ also have the same property as $\Phi_n$
by the same argument: Matter part of $\Delta_n$ are in the form of
(\ref{bstates}), there is no more source for massless components
than $\Delta_0$, and inverses of $L_0$ are well defined.

In general, $A_\mu$
is different from the gauge field $\wt{A}_\mu$
in the effective action except at the leading order, because its
gauge transformation takes different form from the standard one.
They are connected by some field redefinition.
In \cite{est} it has been explained how to compute this field redefinition
order by order.
However, for our solution $A_\mu$ is equal to $\wt{A}_\mu$.
This is because higher order terms of the field redefinition 
contain two or more $A_\mu$ and possibly derivatives, and since $\wt{A}_\mu$
has only one spacetime index, superfluous indices should be contracted with
each other. Therefore higher order terms contain $A_\mu A^\mu$ or
$\p_\mu A^\mu$, which vanish for our solution. Hence our $A_\mu$ is
also an exact solution of the effective action.
This gives another proof of the fact shown in \cite{t}.

%%%%%%%%%%%%%%%%%%%%%%%%%%%%%%%%%%%%%%%%%%%%%%%%%%%%%%%%%%%%%%%%%%%
%%%%%%%%%%%%%%%%%%%%%%%%%%%%%%%%%%%%%%%%%%%%%%%%%%%%%%%%%%%%%%%%%%%
\section{Behavior of massive components}

In this section we investigate how coefficients of massive states in our
solution in the previous section behave by computing those of first and
second massive states coming from $\Phi_1$ and $\Phi_2$, and see more full
order properties suggested by it.

First we compute first massive components. It can be easily seen that
$V_1(k)=c\p X^+\p X^+ e^{ik_iX^i}$ is the only nonzero component and it is
from $\Phi_1$.
Since its conjugate operator is $U_1(k)=
-\frac{2}{(\ap)^2}c\p c\p X_+\p X_+ e^{-ik_iX^i}$,
the component is given by the following:
\bea
& & \int\frac{d^pk}{(2\pi)^p}V_1(k)\vev{U_1(k)|\Phi_1} \nn
& & =\int\frac{d^p k_{(2)}}{(2\pi)^p}\frac{d^p k_{(3)}}{(2\pi)^p}
 V_1(k_{(2)}+k_{(3)}) \nn
& & \times
\left(\frac{4}{3\sqrt{3}}\right)^{2\ap(k_{(2)}^2+k_{(3)}^2
+k_{(2)}\cdot k_{(3)})+1}
 \frac{1}{\ap(k_{(2)}+k_{(3)})^2+1}A_+(k_{(2)})A_+(k_{(3)}).
\eea
We see that the factor $\left(\frac{4}{3\sqrt{3}}\right)^{2\ap
(k_{(2)}^2+k_{(3)}^2+k_{(2)}\cdot k_{(3)})}$ makes the above integral
convergent, since
$\frac{4}{3\sqrt{3}}<1$ and $k_{(2)}^2+k_{(3)}^2+k_{(2)}\cdot k_{(3)}=
(k_{(2)}+\Half k_{(3)})^2+\frac{3}{4}k_{(3)}^2$ becomes large as
$k_{(2)}, k_{(3)}\rightarrow\infty$.

In the case of one-center solution $A_+\propto 1/k^2$,
we plot $F_p(r)$, coordinate expression of the above function,
defined as follows:
\bea
F_p(r) & = & (\ap)^{p-2}\int\frac{d^p k_{(2)}}{(2\pi)^p}
 \frac{d^p k_{(3)}}{(2\pi)^p}e^{i(k_{(2)}+k_{(3)})_ix^i}
 \left(\frac{4}{3\sqrt{3}}\right)^{2\ap(k_{(2)}^2+k_{(3)}^2+k_{(2)}
 \cdot k_{(3)})} \nn
& & \times\frac{1}{\ap(k_{(2)}+k_{(3)})^2+1}
 \frac{1}{k_{(2)}^2}\frac{1}{k_{(3)}^2},
\eea
where $r=\sqrt{(x^i)^2/\ap}$. Figure 1 is the profile of $F_3(r)$.
Note that $F_p$ is real, and 
depends only on $r$ because of the invariance under rotation of $x^i$.
%%%%%%%%%%%%%%%%%%%%%%%%%% figure %%%%%%%%%%%%%%%%%%%%%%%
\begin{figure}[htdp]
\begin{center}
\leavevmode
\epsfbox{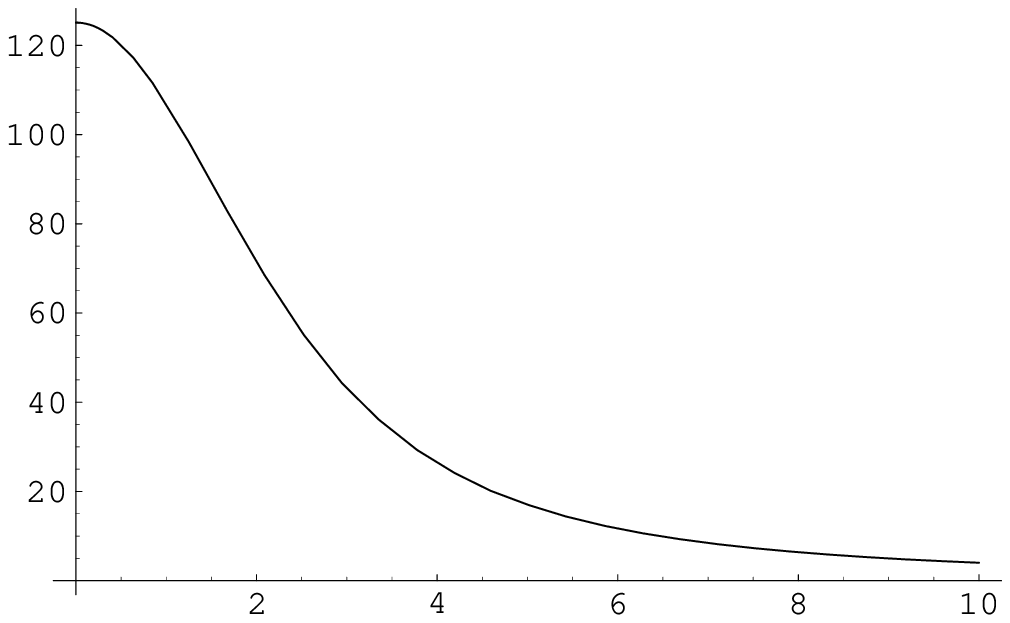}
\caption{$F_3(r)$}  
\label{figure 1}
\end{center}
\end{figure}
%%%%%%%%%%%%%%%%%%%%%%%%%% figure %%%%%%%%%%%%%%%%%%%%%%%

$F_p(r)$ is well-defined everywhere, in particular at $r=0$ unlike
$\Phi_0$. One may wonder why $\Phi_1$, given by the product of $\Phi_0$
which is singular at $r=0$, is smooth. This is because of nonlocality of
the string field product represented by the factor 
$\left(\frac{4}{3\sqrt{3}}\right)^{2\ap(k_{(2)}^2
+k_{(3)}^2+k_{(2)}\cdot k_{(3)})}$.
The nonlocality smears off the singularity. We will see this also happens in
the calculation of higher contribution.

Next we compute a coefficient of a second massive state
$V_2(k)=c\p X^+\p X^+\p X^+ e^{ik_iX^i}$. This is from $\Phi_2$ and other
nonzero second massive states are in $\Phi_1$, which can be computed
similarly to $V_1(k)$. The operator conjugate to $V_2(k)$ is
$U_2(k)=\frac{4}{3(\ap)^3}
c\p c\p X_+\p X_+\p X_+ e^{-ik_iX^i}$. Therefore the component is
\bea
& & \int\frac{d^pk}{(2\pi)^p}V_2(k)\vev{U_2(k)|\Phi_2} \nn
& = & \int\frac{d^pk}{(2\pi)^p}V_2(k)
 \vev{-\frac{b_0}{L_0}U_2(k)\Bigg|\Phi_1\Phi_0+\Phi_0\Phi_1} \nn
& = & \int\frac{d^pk}{(2\pi)^p}V_2(k)
 \left(\frac{4}{3(\ap)^3}\right)\frac{1}{\ap k^2+1}
 \vev{U'_2(k)*\Phi_0+\Phi_0*U'_2(k)|\Phi_1} \nn
& = & \int\frac{d^pk}{(2\pi)^p}V_2(k)
 \left(-\frac{4}{3(\ap)^3}\right)\frac{1}{\ap k^2+1}
 \vev{U'_2(k)*\Phi_0+\Phi_0*U'_2(k)\Bigg|\frac{b_0}{L_0}\Bigg|\Phi_0^2},
\eea
where $U'_2(k)=c\p X_+\p X_+\p X_+ e^{-ik_iX^i}$.
This can be computed in the same way as 4-point amplitudes by noticing
that $\frac{b_0}{L_0}$ is the string field propagator. Coefficients of
higher $\Phi_n$ are also given by $(n+2)$-point off-shell amplitudes.
This fact was pointed out in \cite{kz} in a different context.

Technique for computation of off-shell 4-point amplitudes was developed in
\cite{g,s}. By applying it, we obtain
\bea
& & \vev{U'_2(k)*\Phi_0+\Phi_0*U'_2(k)\Bigg|
 \frac{b_0}{L_0}\Bigg|\Phi_0^2} \nn
& = & \int\frac{d^pk_{(2)}}{(2\pi)^p}\frac{d^pk_{(3)}}{(2\pi)^p}
\frac{d^pk_{(4)}}{(2\pi)^p}
\Bigg[-\frac{3}{8}(\ap)^3(2\pi)^p\delta^p(k_{(2)}+k_{(3)}+k_{(4)}-k)
A_+(k_{(2)})A_+(k_{(3)})A_+(k_{(4)}) \nn
& & \times\int_0^{\sqrt{2}-1}d\alpha
\frac{8\alpha(1-\alpha^2)}{(1+\alpha^2)^3}(\kappa(\alpha))^2 \nn
& & \times
\left(\Half\frac{1+\alpha^2}{1-\alpha^2}\kappa(\alpha)\right)^{\ap(
k^2+k_{(2)}^2+k_{(3)}^2+k_{(4)}^2)}
\left(\frac{2\alpha}{1+\alpha^2}\right)^{2\ap(k_{(3)}+k_{(4)})^2}
\left(\frac{1-\alpha^2}{1+\alpha^2}\right)^{2\ap(k_{(2)}+k_{(3)})^2}
\Bigg],
\label{4int}
\eea
where $\kappa(\alpha)$ is defined in (\ref{kappaa}) in the appendix.

Let us compare the above integral with on-shell Veneziano amplitude.
In the computation of Veneziano amplitude we encounter the following integral:
\beq
\int_0^1dy y^{\ap(k_{(2)}+k_{(3)})^2-2}(1-y)^{\ap(k_{(3)}+ k_{(4)})^2-2}.
\label{ven}
\eeq
This expression is convergent around $y=1$ if $\ap(k_{(3)}+k_{(4)})^2>1$.
Divergence at $\ap(k_{(3)}+k_{(4)})^2=1$ signifies that tachyon mode propagates as an
intermediate state. The integral is not well-defined beyond this point, and
what we usually do is to replace the integral expression by Beta function
which is well-defined except at the poles.

Going back to the expression (\ref{4int}), $1-y$ corresponds to 
$\left(\frac{2\alpha}{1+\alpha^2}\right)^2$, and we can see
(\ref{4int}) does not have
the same problem as (\ref{ven}). This is because $\Phi_1
=-\frac{b_0}{L_0}\Phi_0^2$ does not have
tachyon and massless components as we have shown earlier and these do not
propagate as intermediate states. Therefore we can use the expression 
of moduli integral in (\ref{4int}) for any values of the momenta.

Then another question is the convergence of the integral of the momenta.
Note that $0\leq\left(\frac{2\alpha}{1+\alpha^2}\right)<1$ and 
$0<\left(\frac{1-\alpha^2}{1+\alpha^2}\right)\leq 1$ in the range of $\alpha$.
The equality applies only at the edge of the range. Furthermore
in the appendix we show that 
$0<\left(\Half\frac{1+\alpha^2}{1-\alpha^2}\kappa(\alpha)\right)\leq 1$.
Thus we see that these three factors makes the integral convergent.

The coordinate expression $G_p(r)$ of the above coefficient 
for one-center case, defined as follows, has the profile shown in Figure 2
for $p=3$:
\bea
G_p(r) & = & (\ap)^{3p/2-3}\int
\frac{d^pk_{(2)}}{(2\pi)^p}\frac{d^pk_{(3)}}{(2\pi)^p}\frac{d^pk_{(4)}}{(2\pi)^p} \nn
& & \times\Bigg[e^{i(k_{(2)}+k_{(3)}+k_{(4)})_ix^i}
\frac{1}{\ap(k_{(2)}+k_{(3)}+k_{(4)})^2+2}
\frac{1}{k_{(2)}^2}\frac{1}{k_{(3)}^2}\frac{1}{k_{(4)}^2} \nn
& & \times\int_0^{\sqrt{2}-1}d\alpha
\frac{8\alpha(1-\alpha^2)}{(1+\alpha^2)^3}(\kappa(\alpha))^2
\left(\Half\frac{1+\alpha^2}{1-\alpha^2}\kappa(\alpha)\right)^{\ap(
(k_{(2)}+k_{(3)}+k_{(4)})^2+k_{(2)}^2+k_{(3)}^2+k_{(4)}^2)} \nn
& & \times\left(\frac{2\alpha}{1+\alpha^2}\right)^{2\ap(k_{(3)}+k_{(4)})^2}
\left(\frac{1-\alpha^2}{1+\alpha^2}\right)^{2\ap(k_{(2)}+k_{(3)})^2}
\Bigg].
\eea
%%%%%%%%%%%%%%%%%%%%%%%%%% figure %%%%%%%%%%%%%%%%%%%%%%%
\begin{figure}[htdp]
\begin{center}
\leavevmode
\epsfbox{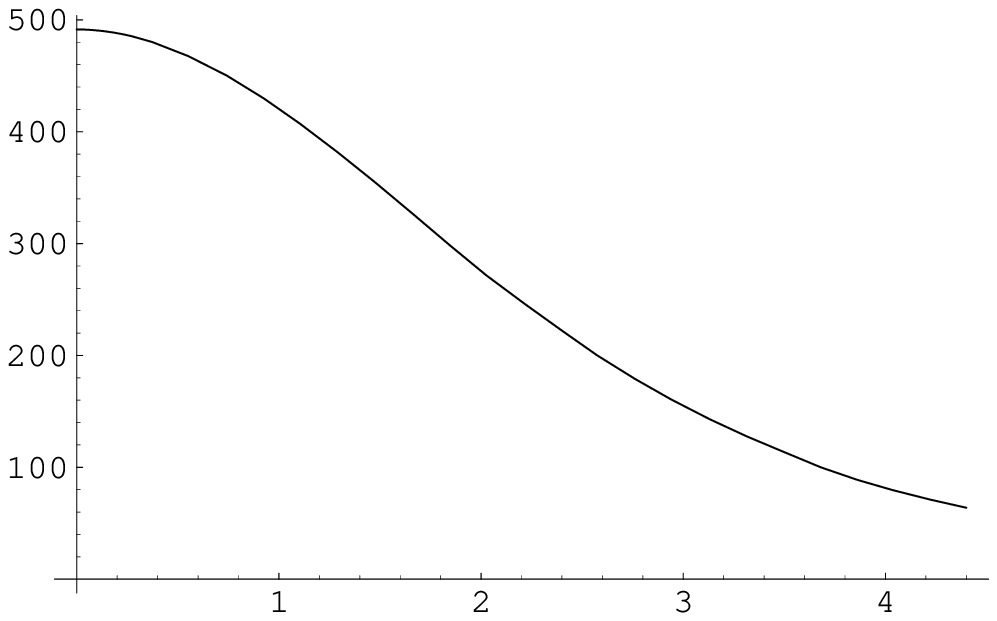}
\caption{$G_3(r)$}  
\label{figure 2}
\end{center}
\end{figure}
%%%%%%%%%%%%%%%%%%%%%%%%%% figure %%%%%%%%%%%%%%%%%%%%%%%

Higher $\Phi_n$ have properties similar to $\Phi_1$ and
$\Phi_2$ i.e. they are related to $(n+2)$-point off-shell amplitudes and 
well-defined, and have smooth profiles. The relation to off-shell amplitudes
implies that integrals of moduli parameters are well-defined at any values
of momenta because $\Phi_n$ do not have tachyon and massless modes, and 
integrals of momenta are convergent even at $r=0$ because
of the nonlocality. Convergent factors come from the following correlator:
\bea
& & \vev{f_1\circ(e^{ik_{(1)}\cdot X})(z_1)
 f_2\circ(e^{ik_{(2)}\cdot X})(z_2)\dots
 f_n\circ(e^{ik_{(n)}\cdot X})(z_n)} \nn 
& & = \prod_i(f_i'(z_i))^{\ap k_{(i)}^2}
 \prod_{i\not\neq j}|f_i(z_i)-f_j(z_j)|^{2\ap k_{(i)}\cdot k_{(j)}}
 (2\pi)^p\delta^p(\sum_i k_{(i)}) \nn
 & & \equiv \exp\left(-\sum_{i,j}a_{ij}k_{(i)}\cdot k_{(j)}\right)
 (2\pi)^p\delta^p(\sum_i k_{(i)}),
\eea
where $f_i(z)$ are conformal transformations appearing in the computation
of off-shell amplitudes. Although we have no 
rigorous proof, we expect that $\sum_{i,j}a_{ij}k_{(i)}\cdot k_{(j)}$
is positive for spatial $k_i$ and works as a convergent factor
for integrals of momenta, because any off-shell string amplitude contains
this factor and it is highly implausible that this is divergent.

The same analysis can be applied to $\Delta_n$:
Although $\Delta_0$ is a sum of delta functions, $\Delta_n$ with $n\geq 1$
are not localized to points and have smooth profiles. This is not
surprising, because the equation of motion is covariant under gauge
transformation, and therefore the source term should also be covariant.
So even if the source term is localized to points in some gauge,
its gauge transformation is not localized due to the nonlocality of the
string star product.

In \cite{cm}, in the free U(1) gauge theory it was shown that the
coefficient in the gauge field $\wt{A}_\mu$ is determined by charge
quantization and the energy around the
singularity $r=0$ is equal to the length times string tension.

In our case the same charge quantization is also applied to $A_\mu$. So we
expect that massive modes do not contribute to the energy. The fact that
massive modes are smooth at $r=0$ also suggests this.
Therefore let us see energy-momentum tensor for our solution.
For definiteness we use the energy-momentum tensor $T_{\mu\nu}$
given in \cite{sen} as Noether current of translation symmetry.
Although this tensor itself is not gauge invariant, total energy and
momentum computed from it are expected to be gauge invariant.
\footnote{I would like to thank A. Sen for clarifying this point.}
This tensor consists of coefficient fields in the string field and
derivatives. Since this has only two spacetime indices $\mu$ and $\nu$,
superfluous indices should be contracted with each other.
We have shown that nonzero component fields have one or more $+$ indices.
If they are contracted with $-$ indices in the derivatives,
we have vanishing contribution because our solution has no $x^\pm$ dependence.
If they are contracted with $-$ indices of other fields, then the $+$ indices
and $-$ indices are paired, and the excess of $+$ indices should be 
$\mu$ and $\nu$. Therefore difference of the number of $+$ and $-$ index
in any nonzero term in $T_{\mu\nu}$ is equal to or less than two.
The only term which satisfies this requirement is $\p^i A_+\p_i A_+$,
and $T_{++}$ is the only nonvanishing component of $T_{\mu\nu}$.

Thus we see that not only the massless modes do not contribute to
$T_{\mu\nu}$, but $T_{\mu\nu}$ is exactly equal to the energy-momentum tensor
of free U(1) gauge theory. Note that the above argument can be applied to
any definition of energy-momentum tensor consisting of two or more
coefficient fields in the string field and derivatives.

One may wonder if the expansion (\ref{phiexp}) is meaningful. 
By the charge quantization $g$ is proportional to the string coupling $g_s$.
In addition, massive modes have no divergent point, and
each coefficient in $\Phi$ receives contribution from only one $\Phi_n$.
We have seen that our solution shares some full order properties with
that of \cite{cm}.
These facts strongly suggest that the expansion (\ref{phiexp}) is
meaningful at least in small $g_s$ region.

%%%%%%%%%%%%%%%%%%%%%%%%%%%%%%%%%%%%%%%%%%%%%%%%%%%%%%%%%%%%%%%%%%%
%%%%%%%%%%%%%%%%%%%%%%%%%%%%%%%%%%%%%%%%%%%%%%%%%%%%%%%%%%%%%%%%%%%
\section{Solution in Superstring Field Theory}

In this section we investigate supersymmetric version of the solution in
the previous sections. We use Berkovits' superstring field theory.
The equation of motion is
\bea
0 & = & \eta_0(e^{-\Phi}Qe^\Phi) \nn
 & = & \sum_{n=0}^\infty\frac{(-1)^n}{(n+1)!}
 \eta_0[\underbrace{\Phi,[\Phi,[\dots,[\Phi}_{n},Q\Phi]]\dots ].
\eea
As in the previous section, we expand $\Phi$ around the solution of the 
linearized equation $\Phi_0$:
\bea
\Phi & = & g\Phi_0+g^2\Phi_1+g^3\Phi_2+\dots, \\
\Phi_0 & = & \int\frac{d^pk}{(2\pi)^p}A_+(k_i)\xi c\psi^+ e^{-\phi}e^{ik_iX^i}.
\eea
$\Phi_n$ satisfy the following equations:
\bea
\Delta_0 & = & \eta_0 Q\Phi_0, \\
\Delta_1 & = & \eta_0\left(Q\Phi_1-\Half[\Phi_0,Q\Phi_0]\right), \\
\Delta_2 & = & \eta_0\left(Q\Phi_2+\frac{1}{6}[\Phi_0,[\Phi_0,Q\Phi_0]]
 -\Half[\Phi_0,Q\Phi_1]-\Half[\Phi_1,Q\Phi_0]\right), \\
 & \vdots & \nn
\Delta_n & = & \eta_0\Bigg(Q\Phi_n \nn
 & & +\sum_{m=1}^n\sum_{\stackrel{
 \mbox{$\scriptstyle n_1, n_2,\dots, n_{m+1}$}}
{\mbox{$\scriptstyle n_1+n_2+\dots+n_{m+1}=n-m$}}}
 \frac{(-1)^m}{(m+1)!}
 [\Phi_{n_1},[\Phi_{n_2},[\dots,[\Phi_{n_m},Q\Phi_{n_{m+1}}]]\dots ]\Bigg), \\
 & \vdots & \nonumber
\eea
where
\beq
\Delta_0=\ap\int\frac{d^pk}{(2\pi)^p}k^2A_+(k_i)c\p c\psi^+ 
e^{-\phi}e^{ik_iX^i}.
\eeq
We impose the gauge fixing conditions $b_0\Phi_n=\wt{G}^-_0\Phi_n=0$, and
for $n\geq 1$ $b_0\Delta_n=0$. This condition, with $\wt{G}^-_0$ defined
as follows\cite{be}, is slightly different from the familiar one
$\xi_0\Phi_n=0$.
\bea
\wt{G}^-_0 & = & \left[Q,\oint\frac{dz}{2\pi i}zb\xi(z)\right] \nn
 & = & \oint\frac{dz}{2\pi i}z(\xi T-\p\xi bc-be^{\phi}G_m
 -\eta e^{2\phi}b\p b),
\eea
where $T$ is the total worldsheet energy momentum tensor, and $G_m$ is
matter part of the worldsheet supercurrent.
This operator is more useful than $\xi_0$ because of the following relations:
\beq
\{\eta_0,\wt{G}^-_0\}=L_0,\quad \{Q,\wt{G}^-_0\}=\{b_0,\wt{G}^-_0\}=0,
\eeq
and therefore $\frac{\wt{G}^-_0}{L_0}$ is the inverse of $\eta_0$ on string
fields annihilated by $\wt{G}^-_0$. Note that $\Phi_0$ obeys
$b_0\Phi_0=\wt{G}^-_0\Phi_0=0$.

Then the equations of motion can be solved order by order:
\bea
\Phi_1 & = & \Half\frac{\wt{G}^-_0}{L_0}\eta_0\frac{b_0}{L_0}
 [\Phi_0,Q\Phi_0], \\
\Phi_2 & = & \frac{\wt{G}^-_0}{L_0}\eta_0\frac{b_0}{L_0}
 \left(-\frac{1}{6}[\Phi_0,[\Phi_0,Q\Phi_0]]
 +\Half[\Phi_0,Q\Phi_1]+\Half[\Phi_1,Q\Phi_0]\right), \\
 & \vdots & \nn
\Phi_n & = & -\frac{\wt{G}^-_0}{L_0}\eta_0\frac{b_0}{L_0}
 \sum_{m=1}^n\sum_{\stackrel{
 \mbox{$\scriptstyle n_1, n_2,\dots, n_{m+1}$}}
{\mbox{$\scriptstyle n_1+n_2+\dots+n_{m+1}=n-m$}}}
 \frac{(-1)^m}{(m+1)!}
 [\Phi_{n_1},[\Phi_{n_2},[\dots,[\Phi_{n_m},Q\Phi_{n_{m+1}}]]\dots ], \\
 & \vdots & \nonumber
\eea
We can see that $\Phi_n$ consists of $Q$, $\eta_0$, $\frac{b_0}{L_0}$,
$\frac{\wt{G}^-_0}{L_0}$ and $(n+1)$ copies of $\Phi_0$.

By plugging the above $\Phi_n$ back into the equations of motion, we obtain
\beq
\Delta_n=\eta_0\frac{b_0}{L_0}Q\sum_{m=1}^n\sum_{\stackrel{
 \mbox{$\scriptstyle n_1, n_2,\dots, n_{m+1}$}}
{\mbox{$\scriptstyle n_1+n_2+\dots+n_{m+1}=n-m$}}}
 \frac{(-1)^m}{(m+1)!}
 [\Phi_{n_1},[\Phi_{n_2},[\dots,[\Phi_{n_m},Q\Phi_{n_{m+1}}]]\dots ].
\eeq
As in the bosonic case, if $\Phi_m$ satisfy equations of motion with 
$\Delta_m=0$ for $m<n$, then $\Delta_n=0$.
To prove this, notice the following identity:
\beq
Q(e^{-\Phi}Qe^{\Phi})+(e^{-\Phi}Qe^{\Phi})^2=0.
\eeq
Therefore
\beq
Q\eta_0(e^{-\Phi}Qe^{\Phi})=
[\eta_0(e^{-\Phi}Qe^{\Phi}), (e^{-\Phi}Qe^{\Phi})].
\eeq
We expand $\Phi$ in $g$ and extract order $g^{n+1}$ contribution of this
equation. From the left hand side,
\beq
Q\eta_0(e^{-\Phi}Qe^{\Phi})|_{g^{n+1}}=Q\eta_0\sum_{m=1}^n\sum_{\stackrel{
 \mbox{$\scriptstyle n_1, n_2,\dots, n_{m+1}$}}
{\mbox{$\scriptstyle n_1+n_2+\dots+n_{m+1}=n-m$}}}
 \frac{(-1)^m}{(m+1)!}
 [\Phi_{n_1},[\Phi_{n_2},[\dots,[\Phi_{n_m},Q\Phi_{n_{m+1}}]]\dots ].
\eeq
Using equations of motion for lower order than $g^{n+1}$, the right hand
side gives
\bea
& & [\eta_0(e^{-\Phi}Qe^{\Phi}), (e^{-\Phi}Qe^{\Phi})]|_{g^{n+1}}= \nn
& & \sum_{l=0}^{n-1}
\Bigg[\Delta_l, \sum_{m=0}^{n-l-1}\sum_{\stackrel{
 \mbox{$\scriptstyle n_1, n_2,\dots, n_{m+1}$}}
{\mbox{$\scriptstyle n_1+n_2+\dots+n_{m+1}=n-l-m-1$}}}
\frac{(-1)^m}{(m+1)!}
 [\Phi_{n_1},[\Phi_{n_2},[\dots,[\Phi_{n_m},Q\Phi_{n_{m+1}}]]\dots ]
\Bigg].
\eea
Therefore
\bea
& & \Delta_n= \nn
& & \frac{b_0}{L_0}\sum_{l=0}^{n-1}
\Bigg[\Delta_l, \sum_{m=0}^{n-l-1}\sum_{\stackrel{
 \mbox{$\scriptstyle n_1, n_2,\dots, n_{m+1}$}}
{\mbox{$\scriptstyle n_1+n_2+\dots+n_{m+1}=n-l-m-1$}}}
\frac{(-1)^m}{(m+1)!}
 [\Phi_{n_1},[\Phi_{n_2},[\dots,[\Phi_{n_m},Q\Phi_{n_{m+1}}]]\dots ]
\Bigg].
\eea
This shows that if $\Delta_m=0$ for $m<n$, then $\Delta_n=0$.

Analogously to the bosonic case,
$\Phi_n$ has no dependence on momenta along $x^\pm$, and
has the following property: $n_+(\omega)-n_-(\omega)=n+1$,
where $\omega$ is any of the vertex operators (or Fock space states)
of which $\Phi_n$ consists,
$n_+(\omega)$ is the number of $\p^mX^+$s and $\p^r\psi^+$
(or $\alpha^+_{-m}$ and $\psi^+_{-r}$) in $\omega$, and 
$n_-(\omega)$ is the number of $\p^mX^-$s and $\p^r\psi^-$
(or $\alpha^-_{-m}$ and $\psi^-_{-r}$) in $\omega$. In addition, 
$(X^I,\psi^I)$ part of $\omega$ is a super-Virasoro descendant of 
the unit operator. In other words, the matter part of $\omega$ is
in the following form:
\bea
& &
\prod_{l=1}^{N_+}\alpha^+_{-p_l}\prod_{l=1}^{M_+}\psi^+_{-q_l}
\prod_{l=1}^{N_-}\alpha^-_{-r_l}\prod_{l=1}^{M_-}\psi^-_{-s_l}
\prod_l L'_{-t_l}\prod_l G'_{-u_l}
\prod_l\alpha^{i_l}_{-v_l}\prod_l\psi^{j_l}_{-w_l}\ket{k_i} \\
& & (N_++M_+-N_--M_-=n+1,\; p_l,r_l,v_l\geq 1,\; t_l\geq 2,
\; q_l,s_l,w_l\geq 1/2,\; u_l\geq 3/2). \nonumber
\eea
where $L'_n$ and $G'_r$ are $(X^I,\psi^I)$ parts of Virasoro operator and 
worldsheet supercharge respectively.

This can be proven by almost the same argument as in the bosonic case.
Here we have new ingredients: $\eta_0$, $Q$ and $\wt{G}^-_0$.
$\eta_0$ does not affect the matter sector. $Q$ and $\wt{G}^-_0$ can
replace $X^\pm$ by $\psi^\pm$ and vice versa, but preserve $n_\pm$. They map
a super-Virasoro descendant of the unit operator to other descendants of it.
$\Delta_n$ also satisfy these properties as can be seen from almost the same
argument.

Therefore this solution has the same properties as in the bosonic case:
each coefficient of Fock space state in the solution $\Phi$ receives
contribution from only one $\Phi_n$. In particular, the coefficient $A_\mu$
of the massless mode $\xi c \psi^+ e^{-\phi}e^{ik_iX^i}$ is never corrected
by higher order contribution. The inverses of $L_0$ in the expression of
$\Phi_n$ with $n\geq 1$ do not cause any problem. $A_\mu$ is equal to the
gauge field in the effective action. This gives another proof of the fact
shown in \cite{t}. Massive modes are convergent even at the singular points
of the massless mode. Energy-momentum tensor as Noether current of
translation symmetry is equal to that of free U(1) gauge theory.

A new property which is not in bosonic theory is supersymmetry. 
Therefore let us investigate supersymmetry of this solution.
Supersymmetry transformation of R-sector string field $\Psi$ is given by
\cite{kt}
\beq
\delta(\eta_0\Psi) = -\eta_0 s(e^{-\Phi}(Q e^{\Phi})),
\eeq
where 
\beq
s=\oint\frac{dz}{2\pi i}e^{i\pi/4}
 \bar{\epsilon}_A\xi(z)e^{-\Half\phi(z)}\Sigma^A(z),
\label{sint}
\eeq
$\bar{\epsilon}_A$ is a constant ten-dimensional Majorana-Weyl spinor,
and $\Sigma^A(z)$ is a spin operator. $e^{-\Half\phi(z)}\Sigma^A(z)$ is 
regarded as Grassmann odd.
The action of $s$ on a string field is defined as the contour integral of
(\ref{sint}) around it.

It is easy to see that the linearized solution $\Phi_0$ is 1/2 supersymmetric
at the linearized level, since on-shell linearized transformation for
massless fields is the same as that of the U(1) gauge theory.
Because of $A_-=A_i=A_I=0$ and $A_+=A_+(k_i)$, the transformation of
gaugino $\psi^A(k)$ is
\beq
\delta \psi^A(k) = ik_i A_+(k_i)(\Gamma^{i+}\epsilon)^A.
\eeq
We see that the unbroken supersymmetry parameter is given by 
$\Gamma^+\epsilon=0$.

In fact, the full solution is also 1/2 supersymmetric with the same unbroken
parameter. This can be shown as follows.
First, notice that when $\Gamma^+\epsilon=0$, $\Phi_0$ satisfies
\beq
s\Phi_0=s\eta_0\Phi_0=sQ\Phi_0=s\eta_0Q\Phi_0=0,
\eeq
and $s$ commutes with $\frac{\wt{G}^-_0}{L_0}$ and $\frac{b_0}{L_0}$.
Then by plugging our solution, $e^{-\Phi}(Q e^{\Phi})$ is expressed by
$\Phi_0$, $\frac{\wt{G}^-_0}{L_0}\eta_0$, $Q$ and $\frac{b_0}{L_0}$.
Using Leibniz rule for $Q$ and $\eta_0$, and $\{Q,\frac{b_0}{L_0}\}=
\{\eta_0,\frac{\wt{G}^-_0}{L_0}\}=1$, we can rewrite $e^{-\Phi}(Q e^{\Phi})$
in such a form that any $Q$ and $\eta_0$ act directly on one of $\Phi_0$.
Since $s$ also satisfies Leibniz rule when it acts on
products of string fields, we can again rewrite $se^{-\Phi}(Q e^{\Phi})$
in such a form that $s$ acts directly on one of $\Phi_0$, $\eta_0\Phi_0$,
$Q\Phi_0$ or $\eta_0Q\Phi_0$. Thus we can see $se^{-\Phi}(Q e^{\Phi})=0$
and therefore $\delta(\eta_0\Psi)=0$.

%%%%%%%%%%%%%%%%%%%%%%%%%%%%%%%%%%%%%%%%%%%%%%%%%%%%%%%%%%%%%%%%%%%
%%%%%%%%%%%%%%%%%%%%%%%%%%%%%%%%%%%%%%%%%%%%%%%%%%%%%%%%%%%%%%%%%%%
\section{Discussion}

We have shown that our solutions have various full order properties in the
sense of $\ap$-expansion. Among them, the fact that massive modes have no
singularity lacks a rigorous proof for third and higher massive states
coming from $\Phi_n$ with $n\geq 3$.
It is desirable to give a proof of it, because this fact is important for
not only our solutions, but also general structure of off-shell amplitudes.

We have constructed higher order source terms for unphysical modes
along with higher order contributions to the solutions, and have seen
that those are not localized to points. This is natural in a sense,
because full order string theory is a nonlocal theory unlike its low
energy effective theory.
Although this is expected not to affect the equation of motion for massless
modes obtained after integrating out all the massive modes, it is better
to give other evidences that our source terms really correspond to
endpoints of fundamental strings.

Readers might wonder why massive modes do not contribute to the
energy-momentum tensor, in spite of the fact that they satisfy Siegel gauge
condition and therefore they are physical excitations. It may be useful to
consider if this fact has any deep meaning for physical properties of
massive modes.

The order by order method employed here can be applied to other systems
e.g. closed string field theory.
It is interesting to construct solutions corresponding to, for example,
macroscopic fundamental string solution or pp-wave solution,
which are also known as $\ap$-exact solutions in supergravity.
We can expect to derive some full order properties of those solutions
by the same method as in this paper.

%%%%%%%%%%%%%%%%%%%%%%%%%%%%%%%%%%%%%%%%%%%%%%%%%%%%%%%%%%%%%%%
\vs{.5cm}
\noindent
{\large\bf Acknowledgments}\\[.2cm]
The author wishes to thank S.\ Iso, Y.\ Okawa and A.\ Sen
for useful discussions, and especially B.\ Zwiebach for reading 
the manuscript and giving helpful comments.
This work is supported in part by funds provided by the U.S. Department of
Energy (D.O.E.) under cooperative research agreement DF-FC02-94ER40818,
and by the Nishina Memorial Foundation.
%%%%%%%%%%%%%%%%%%%%%%%%%%%%%%%%%%%%%%%%%%%%%%%%%%%%%%%%%%%%%%%

\renewcommand{\theequation}{\Alph{section}.\arabic{equation}}
\appendix
\addcontentsline{toc}{section}{Appendix}
%%%%%%%%%%%%%%%%%%%%%%%%%%%%%%%%%%%%%%%%%%%%%%%%%%%%%%%%%%%%%%%
\vs{.5cm}
\noindent
{\Large\bf Appendix}
\section{}
\setcounter{equation}{0}

In this appendix we show that the momentum integral of (\ref{4int})
is convergent, by seeing that the factor
$\Half\frac{1+\alpha^2}{1-\alpha^2}\kappa(\alpha)$ is less than or equal to 1.
First we give the definition of $\kappa(\alpha)$.

4-point amplitudes can be computed by mapping four vertex operators on 
four upper half planes by $w=h_i(Z_i)$, defined as follows,
\bea
h_1(Z)=h_2(Z) & = & {\rm ln}Z-\frac{\tau}{2}, \\
h_3(Z)=h_4(Z) & = & -{\rm ln}Z+\pi i+\frac{\tau}{2},
\eea
and the Giddings map $z=z(w)$ \cite{g}, defined implicitly as follows,
\bea
w & = & \frac{\tau}{2}+N\int_{+0}^zd\zeta
 \frac{\sqrt{\zeta^2+\gamma^2}\sqrt{\zeta^2+\gamma^{-2}}}
 {(\zeta^2-\alpha^2)(\zeta^2-\alpha^{-2})}, \\
N & = & \frac{2\alpha(\alpha^{-2}-\alpha^2)}
 {\sqrt{\alpha^2+\gamma^2}\sqrt{\alpha^2+\gamma^{-2}}},
\eea
to one single upper half plane, on which the four vertex operators are at 
$z=\pm\alpha$ and $z=\pm\alpha^{-1}$.

$\tau$ and $\gamma$ are functions of $\alpha$, and implicitly determined
by the following equations.
\bea
\frac{\pi}{2} & = & N\int_0^\gamma d\zeta\frac{\sqrt{\gamma^2-\zeta^2}
\sqrt{\gamma^{-2}-\zeta^2}}
 {(\zeta^2+\alpha^2)(\zeta^2+\alpha^{-2})}, \\
\tau & = & N\int_\gamma^{\gamma^{-1}}d\zeta\frac{\sqrt{\zeta^2-\gamma^2}
\sqrt{\gamma^{-2}-\zeta^2}}
 {(\zeta^2+\alpha^2)(\zeta^2+\alpha^{-2})}.
\eea
$\gamma$ is a monotonously increasing function, and $0\leq\gamma\leq 1$
as can be seen
from the fact that $z=i\gamma$ and $z=i\gamma^{-1}$ are where two of the
four strings meet, and therefore $z=i\gamma$ is always below $z=i\gamma^{-1}$
on the imaginary axis. $\tau$ is a modulus to be integrated over
$0\leq\tau\leq\infty$ which corresponds to 
$\alpha_0\equiv\sqrt{2}-1\geq\alpha\geq 0$. Near $\alpha=0$, 
$\gamma\sim\sqrt{3}\alpha$, and $\gamma=1$ only at $\alpha=\alpha_0$.
Figure 3 is the profile of $\gamma$.
%%%%%%%%%%%%%%%%%%%%%%%%%% figure %%%%%%%%%%%%%%%%%%%%%%%
\begin{figure}[htdp]
\begin{center}
\leavevmode
\epsfbox{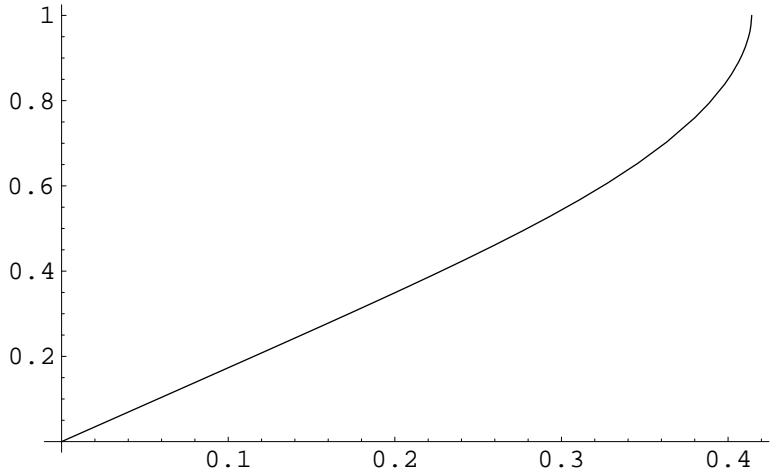}
\caption{$\gamma(\alpha)$}  
\label{figure 3}
\end{center}
\end{figure}
%%%%%%%%%%%%%%%%%%%%%%%%%% figure %%%%%%%%%%%%%%%%%%%%%%%

The above conformal mappings for the vertex operators give the
following factor, which appears in (\ref{4int}):
\beq
(\alpha^{-1}\kappa(\alpha))^{\ap k^2}
(\alpha^{-1}\kappa(\alpha))^{\ap k_{(2)}^2}
(\alpha\kappa(\alpha))^{\ap k_{(3)}^2}
(\alpha\kappa(\alpha))^{\ap k_{(4)}^2},
\eeq
where\footnote{Although this looks different from eq.(3.13) in \cite{s},
this is equal to it as can be seen by partial integration and replacing
${\rm ln}(1-w)$ by $\int^wd\zeta\frac{1}{\zeta-1}$.}
\bea
\kappa(\alpha) & = & \exp(I(\alpha)), \label{kappaa} \\
I(\alpha) & = & \int_0^\alpha d\zeta\Bigg[N
 \frac{\sqrt{\zeta^2+\gamma^2}\sqrt{\zeta^2+\gamma^{-2}}}
 {(\zeta^2-\alpha^2)(\zeta^2-\alpha^{-2})}+\frac{1}{\zeta-\alpha}\Bigg] \nn
& = & \int_0^1 d\zeta\Bigg[N\alpha
 \frac{\sqrt{\alpha^2\zeta^2+\gamma^2}\sqrt{\alpha^2\zeta^2+\gamma^{-2}}}
 {(1-\zeta^2)(1-\alpha^4\zeta^2)}+\frac{1}{\zeta-1}\Bigg].
\eea
The two terms of the integrand are divergent at $\zeta=\alpha$, but their sum
is not. Though it is difficult to perform this integral at generic
$\alpha$, it is possible at the edges of the range of $\alpha$:
\bea
I(0) & = & \frac{8}{3\sqrt{3}}, \\
I(\alpha_0) & = & {\rm ln}\sqrt{2}.
\eea
To show $0<\Half\frac{1+\alpha^2}{1-\alpha^2}\kappa(\alpha)\leq 1$,
we add some extra terms to the integrand which sum up to zero:
\bea
I(\alpha) & = & \int_0^1 d\zeta\Bigg[N\alpha
 \frac{\sqrt{\alpha^2\zeta^2+\gamma^2}\sqrt{\alpha^2\zeta^2+\gamma^{-2}}}
 {(1-\zeta^2)(1-\alpha^4\zeta^2)}
 -2(1-\alpha^2)\frac{1+\alpha^2\zeta^2}{(1-\zeta^2)(1-\alpha^4\zeta^2)}\Bigg]
 \nn
 & & +\int_0^1 d\zeta\Bigg[
 2(1-\alpha^2)\frac{1+\alpha^2\zeta^2}{(1-\zeta^2)(1-\alpha^4\zeta^2)}
 -2(1-\alpha_0^2)\frac{1+\alpha_0^2\zeta^2}{(1-\zeta^2)(1-\alpha_0^4\zeta^2)}
 \Bigg] \nn
 & & +\int_0^1 d\zeta\Bigg[
 2(1-\alpha_0^2)\frac{1+\alpha_0^2\zeta^2}{(1-\zeta^2)(1-\alpha_0^4\zeta^2)}
 +\frac{1}{\zeta-1}
 \Bigg].
\eea
The third integral is equal to $I(\alpha_0)$, and the second integral can 
be explicitly done because the integrand is a rational function: 
\beq
\int_0^1 d\zeta\Bigg[
 2(1-\alpha^2)\frac{1+\alpha^2\zeta^2}{(1-\zeta^2)(1-\alpha^4\zeta^2)}
 -2(1-\alpha_0^2)\frac{1+\alpha_0^2\zeta^2}{(1-\zeta^2)(1-\alpha_0^4\zeta^2)}
 \Bigg]={\rm ln}\left(\sqrt{2}\frac{1-\alpha^2}{1+\alpha^2}\right).
\eeq
The sum of two terms of the integrand in the first integral
is not singular at $\zeta=1$. The final result of this manipulation is
\bea
I(\alpha) & = & {\rm ln}\left(2\frac{1-\alpha^2}{1+\alpha^2}\right) \nn
& & -\frac{2\alpha^2(1-\alpha^2)(1-\gamma^2)}
 {(\alpha^2+\gamma^2)(1+\alpha^2\gamma^2)} \nn
& & \times\int_0^1 d\zeta\Bigg[(1+\alpha^2)
 \sqrt{\frac{(\alpha^2\zeta^2+\gamma^2)(1+\alpha^2\gamma^2\zeta^2)}
 {(\alpha^2+\gamma^2)(1+\alpha^2\gamma^2)}}
 +1+\alpha^2\zeta^2\Bigg]^{-1}.
\eea
Then we obtain the following expression of 
$\Half\frac{1+\alpha^2}{1-\alpha^2}\kappa(\alpha)$:
\bea
\Half\frac{1+\alpha^2}{1-\alpha^2}\kappa(\alpha)
& = & \exp\Bigg(-\frac{2\alpha^2(1-\alpha^2)(1-\gamma^2)}
 {(\alpha^2+\gamma^2)(1+\alpha^2\gamma^2)} \nn
& & \times\int_0^1 d\zeta\Bigg[(1+\alpha^2)
 \sqrt{\frac{(\alpha^2\zeta^2+\gamma^2)(1+\alpha^2\gamma^2\zeta^2)}
 {(\alpha^2+\gamma^2)(1+\alpha^2\gamma^2)}}
 +1+\alpha^2\zeta^2\Bigg]^{-1}\Bigg).
\eea
It is easy to see that the exponent of the right hand side is always
negative, and zero only at $\alpha=\alpha_0$ (where $\gamma=1$).
Thus $0<\Half\frac{1+\alpha^2}{1-\alpha^2}\kappa(\alpha)\leq 1$, and 
the momentum integral of (\ref{4int}) is convergent.
Figure 4 is the profile of 
$\Half\frac{1+\alpha^2}{1-\alpha^2}\kappa(\alpha)$.
%%%%%%%%%%%%%%%%%%%%%%%%%% figure %%%%%%%%%%%%%%%%%%%%%%%
\begin{figure}[htdp]
\begin{center}
\leavevmode
\epsfbox{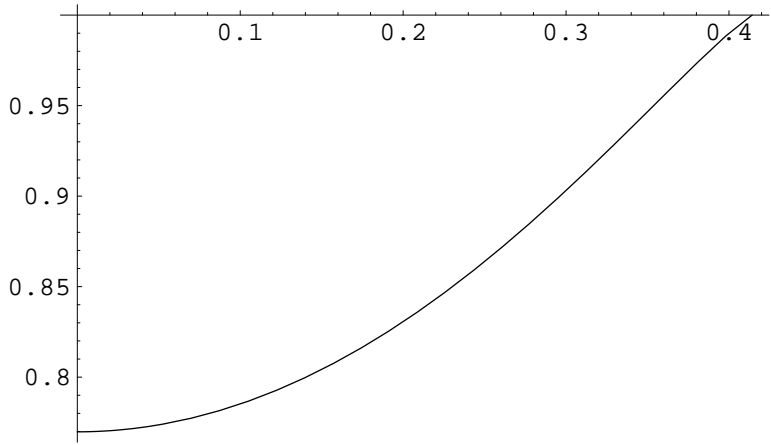}
\caption{$\Half\frac{1+\alpha^2}{1-\alpha^2}\kappa(\alpha)$}  
\label{figure 4}
\end{center}
\end{figure}
%%%%%%%%%%%%%%%%%%%%%%%%%% figure %%%%%%%%%%%%%%%%%%%%%%%

%%%%%%%%%%%% References %%%%%%%%%%%%%%%%%%%%%%%%%
\newcommand{\J}[4]{{\sl #1} {\bf #2} (#3) #4}
\newcommand{\andJ}[3]{{\bf #1} (#2) #3}
\newcommand{\AP}{Ann.\ Phys.\ (N.Y.)}
\newcommand{\MPL}{Mod.\ Phys.\ Lett.}
\newcommand{\NP}{Nucl.\ Phys.}
\newcommand{\PL}{Phys.\ Lett.}
\newcommand{\PR}{Phys.\ Rev.}
\newcommand{\PRL}{Phys.\ Rev.\ Lett.}
\newcommand{\PTP}{Prog.\ Theor.\ Phys.}
\newcommand{\hepth}[1]{{\tt hep-th/#1}}
%%%%%%%%%%%%%%%%%%%%%%%%%%%%%%%%%%%%%%%%%%%%%%%%

\end{document}